# Uniform and Tunable Structural Colors based on Ultrathin Lignin Optical Coatings

Bang An[1,2], Anran Mao[1], Torbjörn Pettersson[1], Wei Li[2], Shouxin Liu[2]*, Anna J. Svagan[1]*, Ravi Shanker [1]*

[1]Royal Institute of Technology (KTH), Dept. of Fibre and Polymer Technology, SE-100 44 Stockholm, Sweden

[2]Key Laboratory of Bio-based Material Science and Technology of Ministry of Education, Northeast Forestry University, Harbin 150040, China.

E-mail: liushouxin@nefu.edu.cn, svagan@kth.se, shanker@kth.se

**ABSTRACT** Structural coloration offers a sustainable and non-fading alternative to conventional pigment- and dye-based colorants. In this study, we present a scalable strategy for generating vivid, tunable structural colors using ultrathin films of industrial kraft lignin. By employing solvent fractionation to isolate a methanol-soluble lignin fraction, followed by spin-coating onto silicon substrates, we achieve highly uniform films with controllable thickness. The observed colors originate from thin-film interference and span the entire visible spectrum by adjusting lignin concentration and spin speed. Spectroscopic ellipsometry and transfer matrix modeling confirm that the interference is modulated by the film thickness. The lignin films exhibit robust mechanical properties, limited angular color shift, and fast, reversible optical response to humidity changes, enabling real-time environmental sensing. This approach circumvents the need for nanoparticle synthesis or self-assembly and highlights the potential of lignin, a widely available and underutilized biopolymer, as a functional photonic material for sustainable optical coatings and colorimetric sensors.

**INTRODUCTION**

Coloration in consumer products is conventionally achieved through the use of dyes and pigments, which produce colors via the selective absorption of specific wavelengths of visible light. These colorants, however, often fade over time and may present environmental and health-related concerns due to their chemical composition and life cycle impacts.[1, 2] A biomimetic and substantially more sustainable alternative is structural colors (SC), wherein color is generated not by absorption but through the interaction of light with organized nanostructures. Structural colors are inherently non-fading, bright, and can be environmentally benign, particularly when fabricated from bio-based materials.

To date, significant efforts have focused on structural color systems derived from cellulose, a renewable and abundant component of plant biomass.[3-5] In contrast, lignin, an almost equally abundant, aromatic biopolymer and major byproduct of the pulping industry, has remained underutilized for such applications. This is because the efficient valorization of lignin at an industrial scale has been hindered by the lignin's complex and varying chemical structure, as well as the presence of impurities, which in part result from the chemical reactions that occur during the pulping process. Lignin is a polyphenolic biomolecule composed of three primary monolignol precursors: *p*-coumaryl alcohol, coniferyl alcohol and sinapyl alcohol.[6] During pulping, the native lignin structure undergoes cleavage of inter-unit linkages and simultaneous formation of new covalent bonds, leading to a highly modified and structurally diverse product. Industrial kraft lignin, which is used in the present study, is produced in large quantities, with approximately 70 million tons of lignin-rich byproducts generated annualy.[7] Unfortunately, it consists of a broad polydisperse mixture of heterogenous molecular structures.[8] To address the molecular heterogeneity, fractionation techniques are commonly employed to isolate more uniform lignin fractions with improved solubility and reproducibility in downstream processing.[9, 10] This step is crucial, as uncontrolled heterogeneity in molecular weight and solubility can negatively impact the mechanical, optical, and processing characteristics of lignin-based materials.[9, 11]

Recent advances in lignin nanotechnology have enabled self-assembled colloidal lignin nanoparticles (LNPs) for structural coloration,[11-16] offering potential for optical coatings and responsive materials. While effective, these systems require additional synthesis and processing steps to achieve a narrow nanoparticle size distribution.[13] Another limitation is that the resulting colors depend on the packing density of the nanoparticles, leading to the formation of either photonic crystals or photonic glasses. Furthermore, some of the self-assembled LNPs structures rely on moisture to exhibit SC,[11, 13] which can shift or be lost upon drying as refractive index contrast decreases , resulting in a darker appearance.[11] In LNPs systems design to exhibit SC under dry conditions,[12, 14] lignin's relatively low hygroscopicity compared to cellulose,[17] and its further tunability through chemical modification,[16] mitigate water uptake to some extent. Despite this, the large supplementary volume fraction of voids (≈ 36 vol% for randomly packed spheres) between LNPs remains a limiting factor, because these voids serve as additional sites for water condensation and water adsorption. The uptake of water into the porous structure then alters the effective refractive index of the materials, in particular at high relative humidities, and influences the response behavior of these nanoparticle-based SC with water content.[16]

In the present work, we introduce an alternative, nanoparticle-free approach for the generation of structural colors based on spin-coated thin films of fractionated kraft lignin. This method simplifies fabrication by avoiding nanoparticle synthesis and self-assembly, while enabling uniform film formation with precisely tunable thickness. By modulating lignin concentration and spin-coating parameters, we demonstrate continuous color tuning across the visible spectrum via thin-film interference. The resulting lignin-based photonic coatings exhibit uniform, vibrant, and environmentally responsive SC. This study thus establishes a scalable and robust platform for utilizing lignin, a low-cost and underexploited biopolymer, in sustainable optical materials.

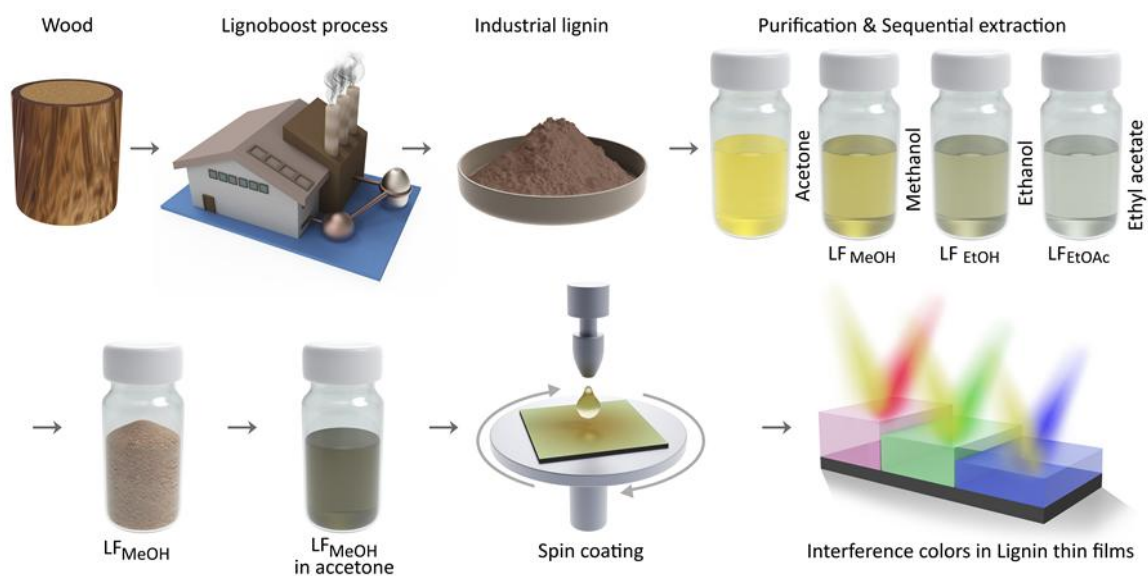

**Figure 1.** Conceptual schematic showing the fabrication of structurally colored lignin-based photonic thin films derived from lignin. The industrial kraft lignin is sequentially fractionated using organic solvents. The methanol-fraction of lignin is spin-coated onto silicon substrates to produce thin films with thickness-dependent structural colors.

**RESULTS AND DISCUSSION**

Industrial softwood kraft lignin (KL), retrieved from black liquor by the LignoBoost technology,[18] was sequentially fractionated using ethyl acetate, ethanol, methanol, and acetone. The methanol-soluble fraction ($\overline{M_w}$ = 2640 g mol$^{-1}$) was selected for this study. We note that this fraction demonstrated excellent solubility in acetone, which was the preferred solvent during the spin coating, due to its lower boiling point than methanol. The lignin solution was spin-coated onto silicon wafers at varying concentrations, resulting in thin films with controlled thickness. The films exhibited vivid structural colors, with the hue directly correlated to film thickness due to thin-film interference effects. Figure 1 presents a schematic overview of the fabrication process and the underlying mechanism of SC generation in lignin thin films.

**Spin-coated thin films based on lignin.** Uniform, structurally colored lignin thin films were achieved by spin coating, a level of control difficult to attain with conventional techniques such as drop casting.[19-21] As depicted in Figure 2a, the rapid evaporation during spin coating produced highly uniform films across millimeter-scale areas, critical for reproducible optical characterization. Figure 2b compares two representative samples: a uniform green (477nm thick) cut from a larger-area deposition (left), and a thinner (300nm thick) purple colored film exhibiting an insignificant radial non-uniformity (right), highlighting the sensitivity of the spin coating process. To further illustrate the morphology, Figure 2c presents a cross-sectional scanning electron microscopic (SEM) image of the green sample with false coloring applied to enhance thickness readability (Figure S1 for the other thicknesses). By systematically tuning both lignin concentration and spin speed (Figure 2d), we achieved a broad palette of vibrant structural colors, from deep blue to vivid red-orange, as evident from the optical micrographs highlighting tunable interference across the visible spectrum. Corresponding digital photographs captured under ambient lighting conditions are provided in Figure S2. This substantial optical tunability surpasses any previous studies on LNPs-based and lignin thin-film-based system, where only a single color (blue) was reported.[22]

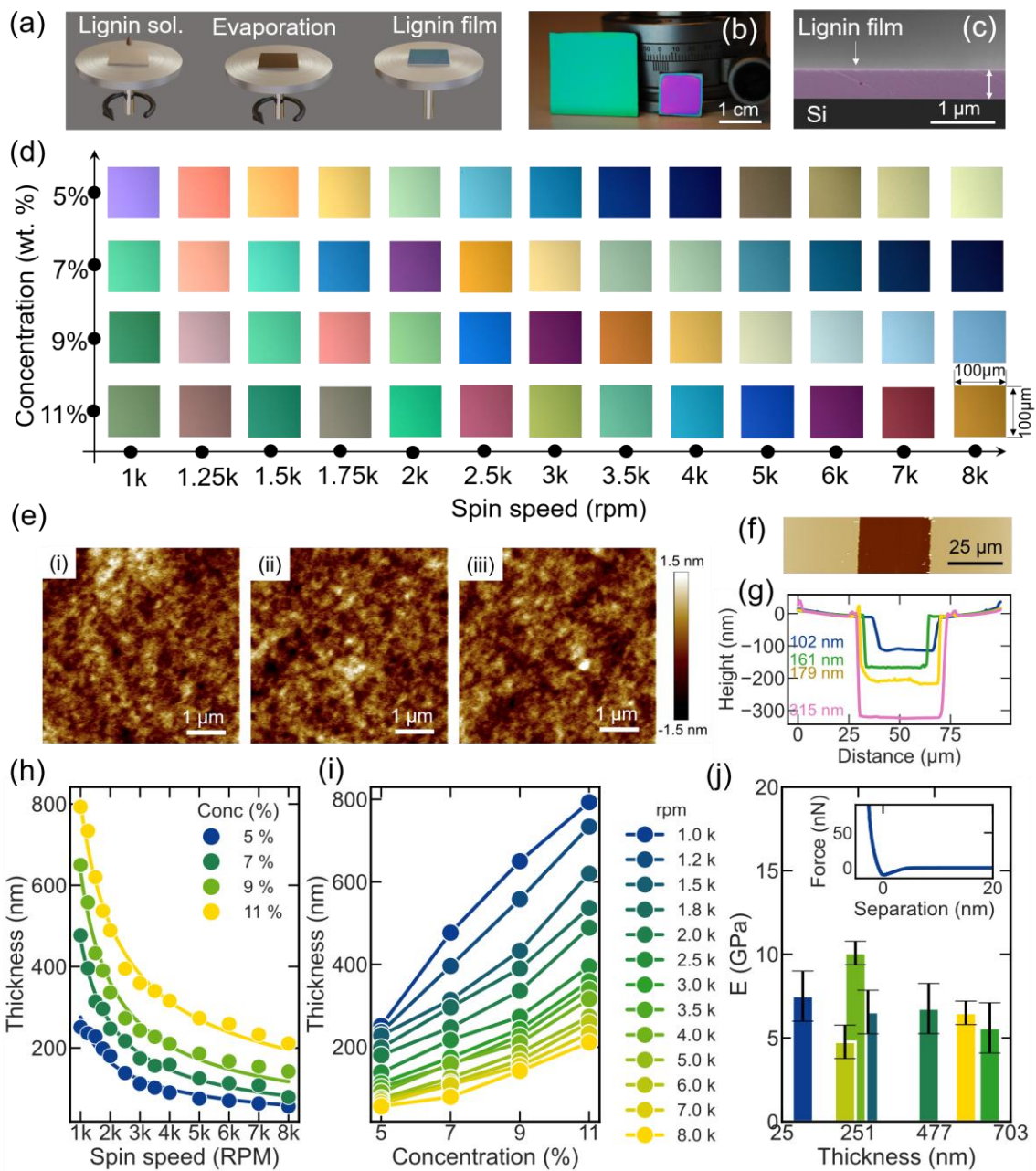

**Figure 2** (a) Schematic of the evolution of spin coating process of lignin thin film. (b) Optical photographs of representative spin-coated samples of two different sizes: a green film (film thickness ~ 477 nm, left) and a purple film (film thickness ~300 nm, right). (c) Cross-sectional SEM image of the green film (film thickness ~ 477 nm) with false-colored thickness. (d) Optical microscopy images showing structural colors as a function of spin speed and lignin concentration. (e) AFM topography height images confirming smooth film surfaces (RMS roughness <1 nm). (f) Step-edge-based AFM thickness measurement of a typical sample and (g) corresponding height profile from selected samples. (h) Lignin film thickness vs. spin speed showing a power-law trend ($n \approx 0.6–0.8$) for the respectively different concentrations. (i) Film thickness increase with lignin concentration at fixed spin speed depending on the spin speed in rpm. (j) The Young's modulus of different lignin films with varying thicknesses, inset showing a representative force vs separation curve for the AFM-indentation measurements.

At lower spin speeds (≤ 3000 rpm), increased rotational velocity sharply reduced film thickness, inducing a pronounced optical blueshift due to reduced interference path length. On the other hand, at fixed spin speeds, increasing lignin concentration consistently yielded thicker films and shifted colors toward longer wavelengths. Because thin-film interference is periodic in thickness ($2nd \approx m\lambda$, where $n$ is the film refractive index, $d$ is the film thickness, and $m$ is the interference order), we observe repeated similar hues in the palette (Figure 2d): SC recur whenever thickness changes by one optical period, $\Delta d = \lambda_{eff}/(2n)$, with $n$ the film refractive index and $\lambda_{eff}$ the effective white-light wavelength; here $\Delta d \approx 170$ nm (e.g, 102 nm and 273 nm). Beyond 3000 rpm, the rate of thickness reduction plateaued, indicating a critical transition where solvent evaporation kinetics, rather than centrifugal forces, predominantly govern the final film thickness and resulting SC. These insights collectively illustrate the robust control achievable over lignin's structural coloration and highlight its exceptional promise as a sustainable, spin-processable optical material.

To elucidate the origin of the observed structural coloration, we quantified film thickness, morphology, and surface roughness using atomic force microscopy (AFM). AFM height images (Figure 2e) showed highly uniform and smooth lignin films (root-mean-square roughness <1 nm), independent of film thickness, confirming the consistency of the spin-coating method. Film thicknesses were also measured with AFM imaging by creating well-defined step edges (Figures 2f and 2g). Figure 2h demonstrates the nonlinear reduction in thickness as spin speed increases. Film thicknesses measured from AFM step-height profiles were fitted to a power-law function ($thickness \propto \omega^{-n}$), to quantitatively assess the dominant film formation mechanisms, yielding exponents $n \approx 0.6-0.8$, indicative of a hybrid thinning regime between evaporation-limited (Meyerhofer, $n$=0.5) and ideal centrifugal thinning (Newtonian, $n=1$) models.[23-25] Notably, at spin speeds below ~3000 rpm, the thickness drops steeply, which in turn correlated with the pronounced color changes observed optically (Figure 2d). On the other hand, at spin speed above ~3000 rpm, when the thickness reduction plateaued, subtle optical shifts were observed as solvent evaporation increasingly dominated over centrifugal spreading. To explicitly illustrate how concentration independently controlled film thickness, the film thickness was plotted as a function of lignin solution concentration at fixed spin speeds (Figure 2i), confirming that thickness can be predictably tuned by varying both concentration and spin speed. Taken together, the observed power-law scaling and linear concentration dependence clearly showed that precise structural color tuning arises from a carefully balanced interplay between polymer viscosity, solvent evaporation kinetics, and centrifugal thinning dynamics, underlining the robustness and versatility of the spin-coating method for fabricating lignin-based photonic thin films.

We then assessed mechanical robustness by measuring film's Young's modulus by AFM indentation as force-displacement measurements. The average Young's modulus values obtained were mainly found in the range of ≈ 4 to 7 GPa (Figure 2j, Figure S3). While consistent with previously reported literature values for lignin,[26, 27] such modulus measurements have not been previously documented for spin-coated lignin films, establishing an important reference point for thin-film applications. Furthermore, the films demonstrate good solvent resistance. The structural colors fully recover after repeated immersion in common solvents such as water and ethanol, followed by nitrogen gas drying, indicating that the film structure is not compromised (Movie S1). To confirm the chemical identity of lignin in the thin films, Fourier-transform infrared spectroscopy (FTIR) was conducted (Figure S4). The FTIR spectra exhibit characteristic bands at ~1510 cm$^{-1}$ (aromatic ring), 1465 cm$^{-1}$, 1270 cm$^{-1}$ (C-O stretch of guaiacyl (G) units), 1140 cm$^{-1}$ (C-H stretch in G units), and 1030 cm$^{-1}$ (C-O, primary alcohol), typical for softwood lignin which is predominantly built from guaiacyl units.[28] The band at ~1720 cm$^{-1}$ corresponds to C=O stretching from carbonyl/carboxyl groups.

We measured the complex refractive index ($n - ik$, where $n$ is the real part and $k$ is the extinction coefficient) by spectroscopic ellipsometry on a representative subset of films. The $n$ and $k$ shown here (Figure 3a-b) are for the 210 ± 2 nm film. The extracted optical constants - $n(\lambda)$ and $k(\lambda)$ exhibit a dispersive profile typical of aromatic polymers. In the visible range (Figure 3a-b), $n(\lambda)$ decreases monotonically with wavelength, $k(\lambda)$ remains low. Over the extended range (200-1500 nm, Figure S5), $n(\lambda)$ shows a peak in the near UV region, and $k(\lambda)$ increases only at shorter wavelength, corresponding to aromatic absorption.[29-31] This wavelength-dependent absorption plays a crucial role in suppressing undesired higher-order resonances, thereby enhancing hue purity and color saturation.[32-34] Using these optical constants, we simulated reflectance spectra of lignin thin films on silicon via transfer matrix modeling (TMM, see SI §1 for details).[35, 36] The calculated reflectance map (Figure 3c) showed thickness-dependent Fabry–Pérot resonances shift with increasing film thickness. Multiple optical modes emerged at higher thicknesses, with their spectral intensities shaped by interference and wavelength-selective absorption.[37-39] Notably, absorption in the blue region helped to suppress off-resonant reflections, resulting in cleaner and more saturated colors.[32] To validate this behavior, we performed micro-reflectance spectroscopy on eight representative lignin films (Figure 3d, Figure S6), selected to span the full visible spectrum from the broader palette in Figure 2d. These samples varied in thickness from ~102 nm to ~793 nm and cover diverse hue categories. The measured reflectance curves show thickness-dependent interference resonances. Thinner film exhibit a single broad peak at shorter wavelength (violet-blue), whereas thicker films support additional resonances and a red shift of the dominant feature. For each sample, the perceived color was computed from the measured spectrum using Commission Internationale de L'Eclairage (CIE) 1931 color matching functions under D65 illumination. Figure 3f presents simulated colors in strong visual agreement with the recorded microscope images (Figure 3e). Quantitatively, the extracted CIE (x, y) chromaticity coordinates from both measured and simulated spectra are in good agreement (Figure 3g).

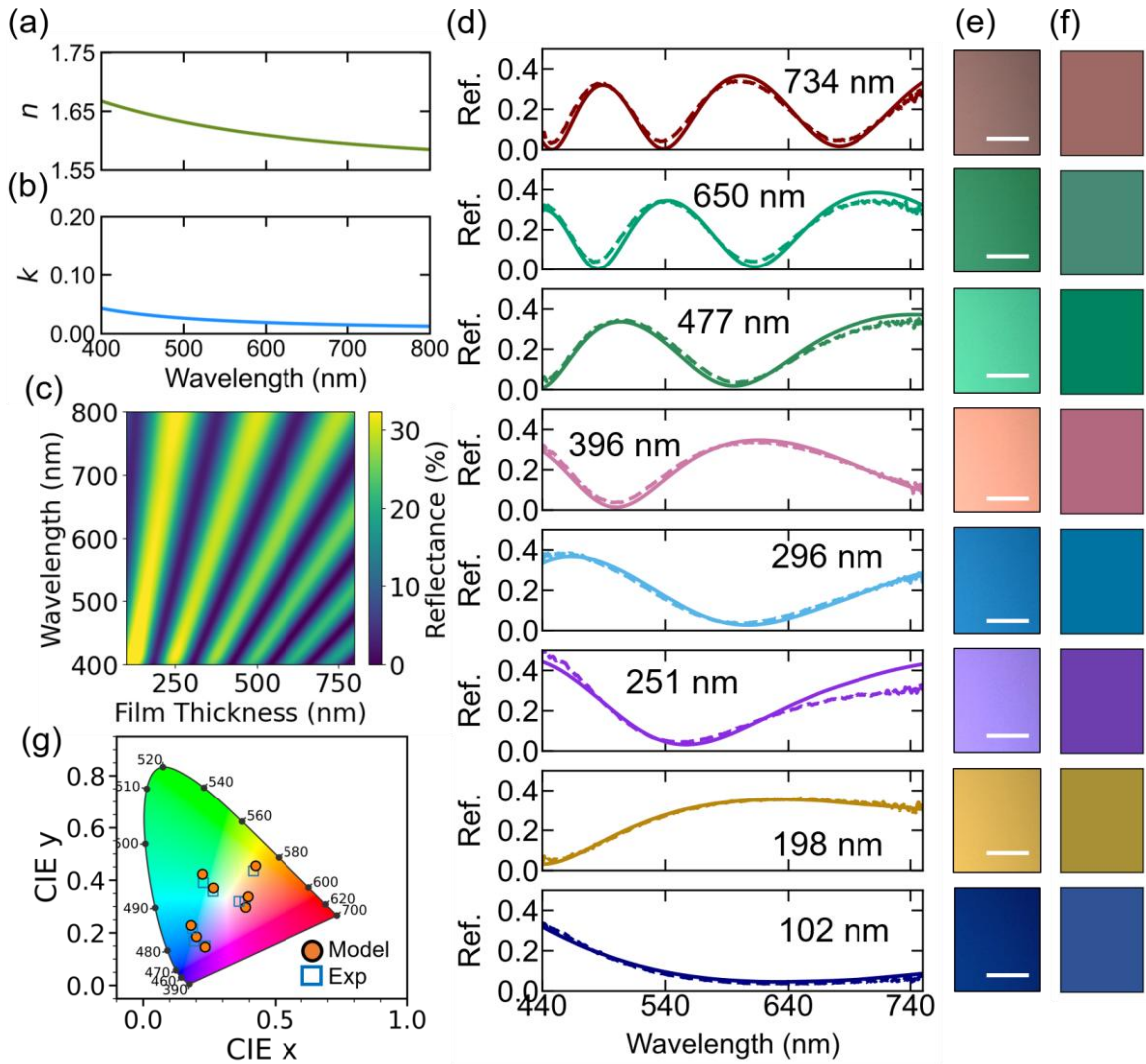

**Figure 3.** Optical properties and structural color generation in spin-coated lignin films. (a) Refractive index $n(\lambda)$ and (b) extinction coefficient $k(\lambda)$ of a ~210 nm lignin thin film obtained by spectroscopic ellipsometry. (c) Simulated reflectance map of lignin films on silicon. (d) Experimental reflectance spectra of eight representative lignin films with varying thickness, selected from the broader palette shown in Figure 2d. (e) Optical microscope images of the same eight samples (scale bar: 50µm). (f) Simulated perceived colors derived from the reflectance spectra in (d), using CIE 1931 color matching functions and D65 illumination. (g) CIE 1931 chromaticity coordinates (x, y) for both measured (orange circles) and simulated (blue squares) reflectance spectra.

**Effect of viewing angle on the perceived structural color**. To evaluate the effect of viewing angle on the perceived structural colors, we conducted angle-resolved reflectance measurements under diffuse illumination (Figure 4a), with samples rotated from 0° to 60°. Figure 4b shows optical photographs of four representative samples (thicknesses: 251, 296, 477, and 734 nm) as a function of viewing angle. All samples exhibit progressive color shifts with increasing angle due to the angle-dependent optical path in the film, for example the 251 nm thick film shifts from magenta towards golden yellow, whereas the 477 nm thick film shifts from green to purple. This trend is captured in Figure 4c, where simulated colors swatches, generated from modeled angle-dependent reflectance spectra for the same film thicknesses, closely track the observations (Figure 4b). Angle-resolved reflectance (Figures 4d–g) shows modest dispersion for $\theta \leq 30°$ ($|\Delta\lambda_{30}|$ = 20-30nm, $\approx 0.8$nm·deg$^{-1}$). For larger angles, both peaks and adjacent minima blue shift, reaching with $|\Delta\lambda_{60}|$ = 66-94 nm ($\approx 1.3$nm·deg$^{-1}$) at $\theta \approx 60°$ relative to normal incidence. To quantify this behavior, we fitted angular positions of the spectral features with a first order interference relation : $\lambda(\theta) = \lambda_0\sqrt{1 - \frac{sin^2\theta}{n^2}}$, where $\lambda(\theta)$ is the resonance wavelength at angle $\theta$, $\lambda_0$ is the normal-incidence wavelength, and $n$ is the effective refractive index of the cavity (Figure 4h).[40] The fitted refractive indices fall within 1.68–1.81, consistent with ellipsometry based estimates. The relatively high refractive index of lignin (~1.7), together with the wavelength-dependent absorption, limits angular dispersion at small tilts and helps maintain hue saturation under oblique views.[41] To quantify color perception, we converted the angle-dependent spectra to chromaticity coordinates in the CIE 1931 color space (Figures 4i–l). These plots reveal gradual coordinate shifts with angle, consistent with the observed color evolution and confirming the predictable angular tunability of the structural colors.

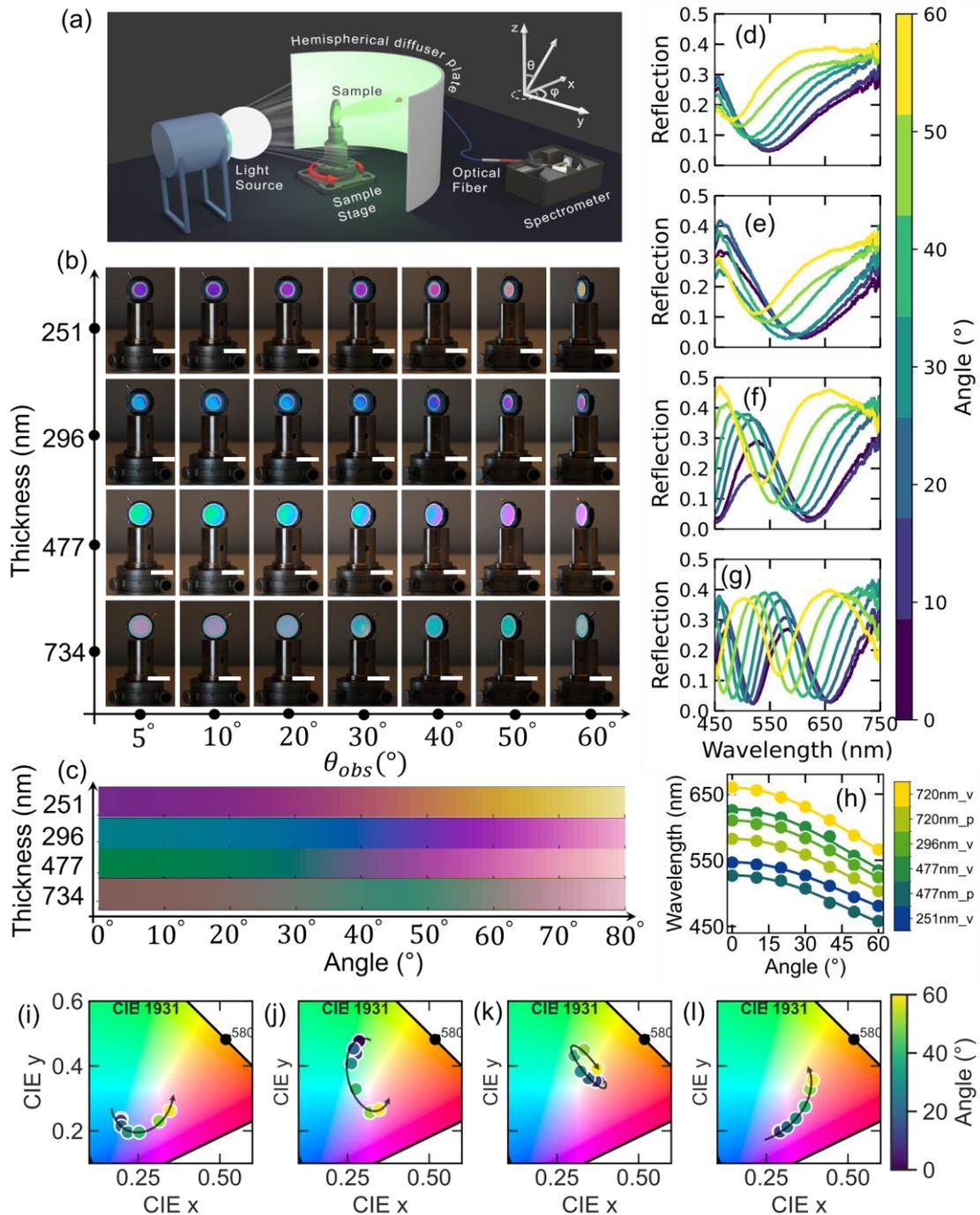

**Figure 4.** Angular dependence of structural colors (a) Schematic of the angle-resolved reflectance measurement setup. (b) Optical photographs of lignin films with varying thicknesses (251–720 nm) at tilt angles from 5° to 60°, showing angle-dependent color shifts. Scale bars: 1.5 cm. (c) Simulated color swatches calculated from modeled spectra for the same samples as indicated from thickness axis. (d–g) Measured reflectance spectra of 251, 296, 477, and 734 nm respectively. (h) Bragg-law fits to extracted peak and valley positions, yielding effective refractive indices ($n$ = 1.68–1.81). (i–l) CIE 1931 chromaticity diagrams showing angular color evolution derived from the spectra in (d–g).

**Humidity responsiveness of the structural color**. Swelling of lignin-thin film at elevated relative humidity (RH) increases the optical thickness $n_{eff}d$, resulting in measurable red-shift of the reflected color. Figure 5a shows a schematic of the RH-induced swelling mechanism. To monitor this effect, a controlled humidity chamber was integrated with a micro-reflectance spectroscopy setup (Figure 5b) to record dynamical optical response of the films under varying RH levels.

We quantify the response as $\Delta\lambda(\text{RH}) = \lambda(\text{RH}) - \lambda(10\%\ \text{RH})$ and plot $\Delta\lambda$ vs RH for four representative films in Figure 5c: blue ($d \approx 296$ nm), green ($d \approx 477$ nm), pink/red ($d \approx 734$ nm), and purple ($d \approx 251$ nm). As RH increases from 20% to 80%, all films show a monotonic red-shift, with $\Delta\lambda$ at 80% RH spanning ~ 30-40 nm. For the blue (d ≈ 296 nm) and green (d ≈ 477 nm) films, the increase in $\Delta\lambda$ over this range is ~24–34 nm, slightly larger than for the pink/red and purple colored samples. The underlying normal-incidence spectra used to extract $\lambda(\text{RH})$ for these same four films with different initial colors and thicknesses are provided in Figure S7. To assess color reversibility, we recorded full-normal incidence reflectance spectra while cycling RH between 20% and 80% over 10 successive cycles (Figure 5d). The applied RH profile is shown in Figure 5e and extracted $\lambda_{peak}$ and $\lambda_{valley}$ traces are plotted in Figure 5f. Both follow the RH profile with minimal hysteresis and return to baseline after each cycle. Within the tested range, no measurable humidity-induced degradation was observed, indicating a reversible and repeatable optical response under humidity cycling.

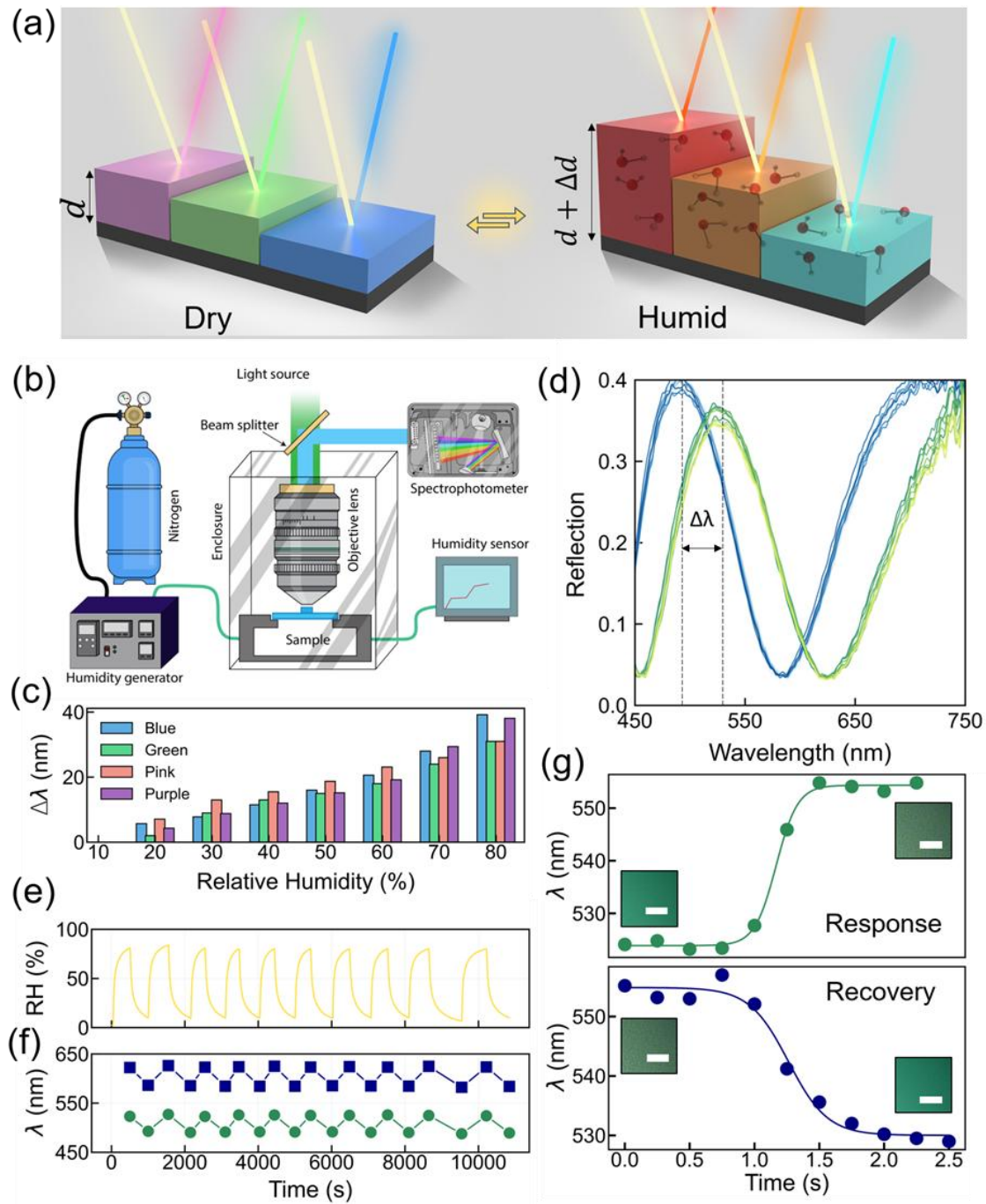

**Figure 5**. Humidity-responsive optical behavior of lignin-based thin films. (a) Schematic illustration of humidity-induced color shift: swelling of the lignin film under high RH increases optical thickness, causing a reversible red-shift for the reflected color. (b) Schematic of the humidity-controlled micro-reflectance measurement setup. (c) Total peak shift ($\Delta\lambda = \lambda(RH) - \lambda(10\% RH)$) as a function of RH for each selected color. (d) Reversibility of reflectance spectra under humidity cycling (20–80% RH, 10 cycles). (e) Measured relative humidity in the sample chamber as a function of time. (f) Tracked peak and trough positions extracted from reflectance spectra. (g) Real-time peak wavelength tracking of a green-reflecting film (~477 nm thick) during one breath cycle, with insets showing microscope images before, and after exposure. Scale bar: 50µm.

To further explore the practical responsiveness and the robustness of the films, we performed a simple breath-induced humidity pulse to a green-color reflecting film of 477 nm thickness. As shown in Figure 5g, a brief exhalation transiently increased the local RH, and the peak wavelength was tracked in real time. The signal shows a rapid red-shift during humidification (top panel) and return blue-shift during drying (bottom panel). The shift magnitude $\Delta\lambda \approx 20$ nm with a response time of ~1 s under ambient conditions (35% RH, 21 °C). Insets of microscope images confirm the reversible color change. The behavior is consistent with a reversible change in optical thickness $n_{eff}d$ due to water uptake. Comparable behavior responses were observed for other film thickness, with $\Delta\lambda \approx 30$-50 nm (Figure S6) under the same perturbation. The films also showed robust, reversible optical response, retaining their color after repeated immersion–dry cycles in water and ethanol (Movie S2-S5).

The dynamic vapor sorption results in Figure 6a, further demonstrated the advantage of the present thin films over nanoparticle-based thin films, photonic crystals, and photonic glasses. Both isothermal sorption curves exhibit type II behavior, i.e., rapid uptake at low RH followed by gradual multilayer adsorption at higher RH. Across all relative humidity (RH) levels, the nanoparticle-based materials sorbed significantly more of water, with the difference being especially pronounced above ≈ 50% RH. In this high-humidity range, the DVS curve for the lignin nanoparticle-based material displayed a much steeper slope, consistent with capillary condensation in pores and cavities. Notably, although that the lignin used in the nano-particles system was acetylated, reducing its intrinsic hygroscopicity, the presence of pores still led to substantial moisture sorption.[42, 43] From the isothermal desorption curve (45-95% RH) of the current solid lignin films, we estimate a pore volume of approximately 5 vol%,[44, 45] indicating a significantly lower porosity. This reduced water uptake is beneficial, as it minimizes RH-induced changes in the effective refractive index, thereby improving color stability under fluctuating environmental conditions.

Finally, we assessed the color range achievable with our spin-coated, thickness-tuned lignin films. By changing only the film thickness, we can produce a wide range of vivid structural colors across the visible color space (Figure 6b). Using our spin-coated film thickness tuning, we achieved 56.3 % coverage of the sRGB gamut—substantially higher than the 17–28 % typical of nanoparticle approaches while also enabling a uniform, continuous range of intermediate hues.[11, 12, 42, 46, 47] While the single-layer design provided an extensive color range, some limitations remained especially for highly saturated red hues due to refractive index dispersion and the thickness limitation achievable by spin-coating. These results open opportunities for constructing multilayer or full-cavity architectures to achieve colors not accessible in the current geometry.

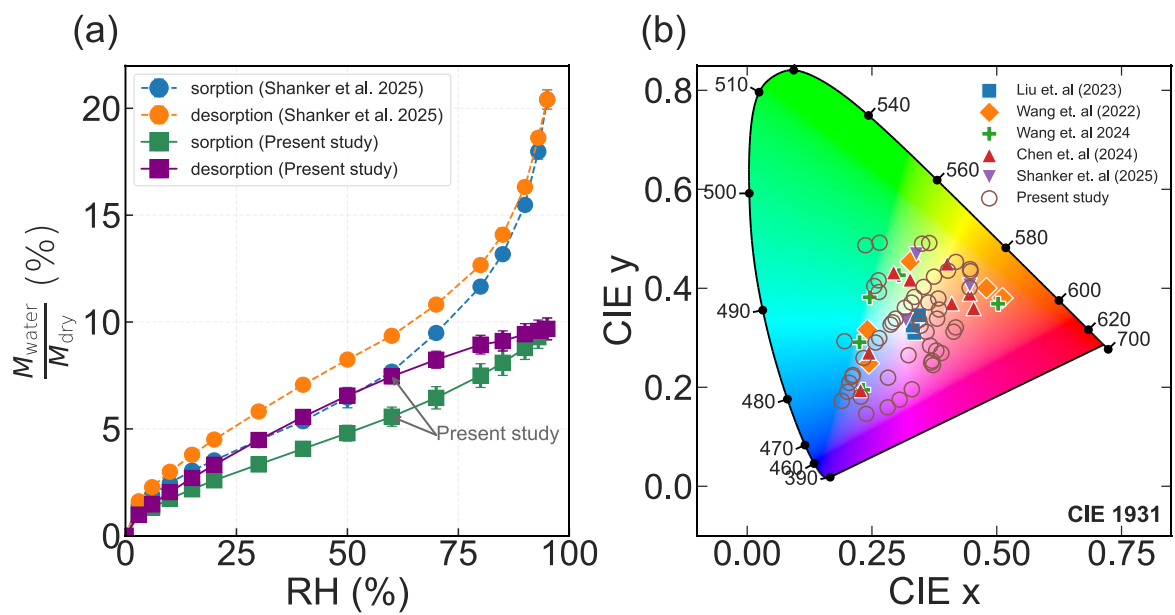

**Figure 6**. Water sorption and chromaticity mapping of lignin-based thin films. (a) DVS isotherms showing lower water uptake in solid lignin films (present study) vs. nanoparticle-based films (Shanker et al. 2025[42]), where $M_{water}$ is the water mass and $M_{dry}$ is the dry sample mass. (b) CIE1931 chromaticity diagram comparing thickness-tuned lignin films and other reported lignin-based from the recent literature.[11, 12, 42, 46, 47]

## Conclusion

This study establishes a robust strategy for generating structurally colored films from spin-coated lignin, a naturally abundant and UV-absorbing biopolymer. By tuning concentration and spin speed, we access a wide and reproducible color palette that spans from violet to red with chromaticity values broadly distributed across the visible spectrum. Optical characterization reveals that this tunability arises from asymmetric Fabry–Pérot resonances within the single-layer films, governed by interference and modulated by intrinsic absorption. Transfer matrix modeling using ellipsometry-derived optical constants reproduces both the spectral features and perceived colors with high accuracy. Lignin films exhibited angular dependence, yet the visual appearance remained similar within a viewing angle range of 0° to approximately 30°, enabled by the absorbing cavity design and diffuse illumination. Furthermore, the colors are not static: the films show an fast, reversible optical response to humidity, with sub-second switching due to transient swelling. This combination of sustainable material sourcing, simple fabrication, broad color tunability, and fast environmental responsiveness highlights lignin's untapped potential as a functional photonic material for future optical coatings, displays, and sensors.

## Acknowledgements

Financial support was granted by Olle Byggmästare (Grant No. 194-0679). Bang An is grateful for the support of the CSC-KTH program (Grant No. 202406600003). N. Krivánková is acknowledged for helping in size-exclusion chromatography measurement, Dr. D. Mahlin is acknowledged for help with DVS and Dr. Y. Cui is acknowledged for lignin fractionation.

## Conflict of Interest

The authors declare no conflict of interest.

# Experimental Section

**Materials**

The softwood Kraft Lignin (KL) was obtained from the Lignoboost process and originated from Stora Enso (Finland). The KL was washed (acidic water, pH 2) and sequentially solvent fractionated following a previously developed protocol.[10, 48] Briefly, the different fractions are obtained by subsequent dissolution in first ethyl acetate, separating out the insoluble fraction that was subsequently dissolved in ethanol. The non-dissolved part was subsequently dissolved in methanol, and the procedure was then repeated with acetone. In the present study, we used the lignin that was solubilized in methanol, which we observed is also soluble in acetone. We prepared lignin solutions by dissolving the aforementioned lignin in 96% acetone at varying ratios, followed by overnight magnetic stirring. The solution was subsequently filtered twice using a PTFE syringe filter with a pore size of 0.22 μm. This process yielded solutions with lignin concentrations of 5 wt%, 7 wt%, 9 wt%, and 11 wt%. The final concentrations of the solutions were verified through TGA analysis. Specifically, we heated the known mass of lignin solution at a rate of 0.5 °C per minute until it reached 800 °C. We took the mass at 120 °C as the mass of dry lignin, and calculated the content of lignin in the lignin solution.

**Spin coating assembly of lignin thin film on silicon wafer**

Circular silicon wafers were purchased from Alpha Nanotech and Shunsheng Electronics (Zhejiang, China). Acetone was purchased from Avantor, and ethanol was purchased from Sigma-Aldrich. All chemicals and materials were used as received without further purification. Silicon wafers were cleaned using a multi-step process. Initially, the wafers were successively ultrasonically cleaned with acetone and ethanol for 5 minutes each. The acetone cleaning is to remove the organic contaminants on the surface of the silicon wafer, while the ethanol cleaning is to remove the residual acetone. They were then washed with Milli-Q water for another 10 minutes to ensure the removal of organic solvents from the surface of the silicon wafer, followed by a 5-minute rinse with acetone. After being dried by nitrogen blowing, it is then treated with a plasma cleaning machine to make it hydrophilic (PDC-002, Harrick scientific corp., USA). Lignin thin films were prepared by a spin-coating technique (EZ4 model from LEBO Science, China). Lignin films were prepared by depositing 0.5 mL of lignin–acetone solution onto the center of a silicon wafer mounted on the spin coater. The spin speeds used were 1000, 1250, 1500, 1750, 2000, 2500, 3000, 3500, 4000, 5000, 6000, 7000 and 8000 rpm. For samples spun below 4000 rpm, the coater was accelerated at 1000 rpm s$^{-1}$ to the target speed, held for 120 s, then ramped to 4000 rpm and maintained for an additional 30 s to remove excess solvent from the edges. For samples at or above 4000 rpm, the speed was held constant for 120 s after reaching the target, using the same acceleration rate of 1000 rpm s$^{-1}$.

**Characterization**

**Scanning electron microscopy** (SEM), was used to examine the surface and cross-sectional morphology of the lignin thin films. Micrographs were acquired using a Hitachi S-4800 SEM (Japan) at an accelerating voltage of 1 kV. Samples were taken from spin-coated lignin films on 1 × 1 cm silicon wafers. Selected samples, prepared on silicon wafers, were subsequently cut into smaller pieces for top-surface imaging. For cross-sectional SEM imaging, the film and substrate were fractured to obtain a clean break. All samples were dried at room temperature and then sputter coated with a 2-3 nm layer of platinum-palladium (Pt/Pd) coating (Cressington scientific instrument, Model no - 208HRD sputter coater, UK) prior to imaging.

**Fourier-transform infrared** (FT-IR) spectra were collected using a Perkin-Elmer Spectrum 2000 FT-IR equipped with a MKII Golden Gate single reflection ATR system, recording spectra in the range of 600 to 4000 cm$^{-1}$ with an average of 32 scans at a resolution of 4.0 cm$^{-1}$.

**Spectroscopic ellipsometry** was used to determine the complex refractive index (n, k) of lignin thin films deposited on silicon substrates. Measurements were performed using a ME-L Ellipsometer (Eoptics technology system, China) at a fixed angle of incidence of 65°, over the spectral range of 200–1100 nm. Each sample was measured twice. Film thickness was obtained by fitting the ellipsometric data using a Cauchy distribution model, and cross-validated with independent thickness measurements from AFM.

**Molecular weight characterization of lignin was performed using size-exclusion chromatography (SEC)** of lignin was performed using size-exclusion chromatography (SEC). Lignin samples were dissolved at a concentration of 4 mg/mL in DMSO containing 0.5 wt% lithium bromide (LiBr) and filtered through a 0.45 μm PTFE syringe filter prior to injection. The SEC system consisted of a PSS SECcurity® setup equipped with a PSS SECcurity degasser, an Agilent 1260 Isopump G1310B, an Agilent 1260 ALS G1329B autosampler, and a PSS SECcurity TCC7000 column oven maintained at 60 °C. The separation was carried out on PSS Gram analytical columns (1000 Å and 10,000 Å, 300 mm × 8.0 mm, 10 μm particle size), preceded by a PSS Gram precolumn (8 mm × 8.0 mm, 10 μm). The system included a variable wavelength UV detector (Agilent 1260 VWD VL G1314B, set at 280 nm), a multi-angle light scattering (MALLS) detector (PSS SLD7100, λ = 627 nm, 7 angles), and a differential refractive index (RI) detector (Agilent RID G1362A), operated at 40 °C. The mobile phase was DMSO with 0.5 wt% LiBr, delivered at a flow rate of 0.5 mL/min. Injection volume was 100 μL, and total analysis time per sample was 65 minutes. Molecular weight averages (Mn, Mw) and dispersity (Đ) were determined using MALLS and RI data (Figure S9), with system calibration based on Pullulan standards covering a wide molar mass range (342–708,000 Da).

**AFM** measurements were used for both topographical imaging and mechanical properties. To determine film thickness, sharp edges were created on the samples by gentle scratching with a sharp tweezer, and height profiles were recorded in tapping mode. Nanomechanical properties were measured by force spectroscopy indentation on the same setup using (Multimode 8, Bruker) equipped with an EV scanner and an RTESPA-150 cantilever (nominal spring constant ~5 N/m, tip radius ~8 nm). The cantilever type was selected to minimize detector non-linearity.[49] The inverse optical lever sensitivity was obtained on a clean silica wafer, and the spring constant was calibrated via the built-in thermal tune function.[50] Indentation was performed at multiple locations using a 500 nm ramp size, 1 μm/s approach speed, and a trigger force of ~600 nN, resulting in indentation depths between 5–40 nm depending on sample stiffness. Force–distance curves were converted into force–indentation data, and Young's modulus was extracted by fitting the linearized Hertz model (force range: 10-60 nN, Poisson's ratio: 0.3, tip radius: 20 nm)[51, 52] using AFM Force IT v3 software (ForceIT, Sweden).[53]

**Optical measurements** were performed using a custom-built micro-reflectance setup integrated with a controlled humidity chamber. The system enabled simultaneous acquisition of optical microscope images and reflectance spectra under variable humidity conditions. Relative humidity (RH) was adjusted by supplying pressurized nitrogen through a P-series humidifier (Cellkraft AB, Stockholm, Sweden), and measurements were conducted within a sealed chamber. A 50 W halogen/LED light source illuminated the sample, which was observed through a Motic BA310Pol microscope equipped with a Moticam ProS5 Lite camera. Reflected light was collected from the trinocular port via an optical fiber and directed to a CCS200M spectrometer (Thorlabs). Spectra were recorded using a 40× objective (NA = 0.65), and all reflectance values were normalized against a protected silver mirror (ME1-P01, Thorlabs). A dark spectrum (0% reflectance) was obtained by blocking the illumination path. The same setup was used to study humidity-triggered optical changes, including dynamic response tests and reversibility over multiple RH cycles (typically between 20% and 80% RH). Before each humidity cycle, the chamber was first purged to ≈5 % RH (well below our nominal 20 % "low" set-point) to ensure a fully dried baseline and stable starting condition. In addition, breathing tests were conducted by exposing the sample to a single breath and tracking the real-time reflectance shift, enabling assessment of response time and recovery behavior under a practical humidity stimulus. Angle-dependent reflectance and color appearance were measured using a custom setup with a 50 W halogen/LED light source and a rotational stage (Thorlabs). The sample was rotated from 0° to +60°, while illumination and detection remained fixed. Reflected light was directed to a Thorlabs CCS200M spectrometer via optical fiber. All spectra were normalized to a silver mirror (ME1-P01). Photographs at each angle were captured to compare spectral shifts with visible color changes.

**Simulating Color Changes in Lignin Films**: Based on the measured complex refractive index of lignin, we simulated angle-resolved reflectance spectra using the transfer matrix method (Supporting info Section §1).[36, 54] These spectra were then converted into RGB color values, which allowed us to visualize how the perceived structural color switches with both film thickness and viewing angle. In the Methods we compute gamut coverage by first extracting the CIE 1931 (x,y) coordinates for each film thickness and all literature datasets, then constructing the convex hull of those points using the SciPy ConvexHull routine.[55] The hull area is evaluated with a simple polygon-area formula and normalized to the area of the sRGB triangle (vertices at (0.64, 0.33), (0.30, 0.60), (0.15, 0.06)) to yield % sRGB coverage. For full coverage, the same hull area is instead divided by the area of the entire spectral-locus horseshoe (sampled at 1 nm from 380–780 nm). All calculations and plotting were performed in Python with numpy, scipy, and the colour library.

**Dynamic vapor sorption measurements:** Water sorption and desorption measurements were performed using a DVS Discovery SA (TA Instruments) at 21 °C (0 – 95% RH). Sample size was 4 mg. Equilibrium for each RH-step was considered reached when dm/dt < 0.002 % for 10 min, or a maximum of 360 min per step. The gas flow rate was 200 mL min$^{-1}$, and nitrogen was used as the carrier gas. The pore volume fraction between 45-95 %RH was estimated from the desorption curve. The mass of the desorbed water in this interval was divided by the density of water ($\rho_{water}$ = 0.997 g cm$^{-3}$) to calculated the corresponding water volume, which was taken as an estimate of the pore volume. To determine the pore volume fraction, the calculated pore volume was divided by the total material volume, which was defined as the sum of the dry material volume (obtained by dividing the dry mass by the solid density of lignin, $\rho_{lignin} \approx 1.6$ g cm$^{-3}$) and the volume of the total water content (total mass of sorbed water between 0% - 95% RH, divided by the density of water).

## REFERENCES

(1) Chung, K.-T. Azo dyes and human health: A review. *Journal of Environmental Science and Health, Part C* **2016**, *34* (4), 233-261. DOI: 10.1080/10590501.2016.1236602.
(2) Turner, A. Cadmium pigments in consumer products and their health risks. *Sci Total Environ* **2019**, *657*, 1409-1418. DOI: https://doi.org/10.1016/j.scitotenv.2018.12.096.
(3) Anusuyadevi, P. R.; Shanker, R.; Cui, Y.; Riazanova, A. V.; Jarn, M.; Jonsson, M. P.; Svagan, A. J. Photoresponsive and Polarization-Sensitive Structural Colors from Cellulose/Liquid Crystal Nanophotonic Structures. *Adv Mater* **2021**, *33* (36), e2101519. DOI: 10.1002/adma.202101519.
(4) Parker, R. M.; Guidetti, G.; Williams, C. A.; Zhao, T.; Narkevicius, A.; Vignolini, S.; Frka-Petesic, B. The Self-Assembly of Cellulose Nanocrystals: Hierarchical Design of Visual Appearance. *Advanced Materials* **2018**, *30* (19), 1704477. DOI: https://doi.org/10.1002/adma.201704477.
(5) Phoungtawee, P.; Sudyoadsuk, T.; Pettersson, T.; Crespy, D.; Svagan, A. J.; Shanker, R. Spatiotemporal Retention of Structural Color and Induced Stiffening in Crosslinked Hydroxypropyl Cellulose Beads. *Macromolecular Rapid Communications* **2025**, *46* (5), 2400755. DOI: https://doi.org/10.1002/marc.202400755.
(6) Whetten, R. W.; MacKay, J. J.; Sederoff, R. R. RECENT ADVANCES IN UNDERSTANDING LIGNIN BIOSYNTHESIS. *Annual Review of Plant Biology* **1998**, *49* (Volume 49, 1998), 585-609. DOI: https://doi.org/10.1146/annurev.arplant.49.1.585.
(7) Poveda-Giraldo, J. A.; Solarte-Toro, J. C.; Alzate, C. A. C. The potential use of lignin as a platform product in biorefineries: A review. *Renewable and Sustainable Energy Reviews* **2021**, *138*, 110688.
(8) Crestini, C.; Lange, H.; Sette, M.; Argyropoulos, D. S. On the structure of softwood kraft lignin. *Green Chem* **2017**, *19* (17), 4104-4121. DOI: 10.1039/c7gc01812f.

(9) Gioia, C.; Lo Re, G.; Lawoko, M.; Berglund, L. Tunable Thermosetting Epoxies Based on Fractionated and Well-Characterized Lignins. *J Am Chem Soc* **2018**, *140* (11), 4054-4061. DOI: 10.1021/jacs.7b13620.

(10) Cui, Y.; Lawoko, M.; Svagan, A. J. High Value Use of Technical Lignin. Fractionated Lignin Enables Facile Synthesis of Microcapsules with Various Shapes: Hemisphere, Bowl, Mini-tablets, or Spheres with Single Holes. *ACS Sustainable Chemistry & Engineering* **2020**, *8* (35), 13282-13291. DOI: 10.1021/acssuschemeng.0c03521.

(11) Wang, J.; Chen, W.; Yang, D.; Fang, Z.; Liu, W.; Xiang, T.; Qiu, X. Photonic lignin with tunable and stimuli-responsive structural color. *ACS nano* **2022**, *16* (12), 20705-20713.

(12) Liu, J.; Nero, M.; Jansson, K.; Willhammar, T.; Sipponen, M. H. Photonic crystals with rainbow colors by centrifugation-assisted assembly of colloidal lignin nanoparticles. *Nature Communications* **2023**, *14* (1), 3099.

(13) Chen, J.; Tian, J.; Feng, N.; Ning, L.; Wang, D.; Zhao, B.; Guo, T.; Song, J.; Rojas, O. J. Monodispersed Renewable Particles by Cascade and Density Gradient Size Fractionation to Advance Lignin Nanotechnologies. *Small* **2024**, 2309756.

(14) Wang, J.; Chen, W.; Yang, D.; Fang, Z.; Liu, W.; Xiang, T.; Qiu, X. Monodispersed Lignin Colloidal Spheres with Tailorable Sizes for Bio-Photonic Materials. *Small* **2022**, *18* (19), 2200671.

(15) Henn, K. A.; Babaeipour, S.; Forssell, S.; Nousiainen, P.; Meinander, K.; Oinas, P.; Österberg, M. Transparent lignin nanoparticles for superhydrophilic antifogging coatings and photonic films. *Chemical Engineering Journal* **2023**, *475*, 145965.

(16) Shanker, R. Structurally Colored Thin Films based on Acetylated Lignin Nanoparticles. *ACS Nano* **2025**. DOI: 10.1021/acsnano.4c16679.

(17) Hou, S.; Wang, J.; Yin, F.; Qi, C.; Mu, J. Moisture sorption isotherms and hysteresis of cellulose, hemicelluloses and lignin isolated from birch wood and their effects on wood hygroscopicity. *Wood Sci Technol* **2022**, *56* (4), 1087-1102. DOI: 10.1007/s00226-022-01393-y.

(18) Tomani, P. The lignoboost process. *Cellulose Chemistry & Technology* **2010**, *44* (1), 53.

(19) Nabhan, M. A.; Silvera Batista, C. A.; Cliffel, D. E.; Jennings, G. K. Spin coating photoactive photosystem I–PEDOT: PSS composite films. *ACS Applied Polymer Materials* **2023**, *5* (5), 3278-3288.

(20) Hall, D. B.; Underhill, P.; Torkelson, J. M. Spin coating of thin and ultrathin polymer films. *Polymer Engineering & Science* **1998**, *38* (12), 2039-2045.

(21) Extrand, C. W. Spin coating of very thin polymer films. *Polymer Engineering & Science* **1994**, *34* (5), 390-394.

(22) Norgren, M.; Notley, S. M.; Majtnerova, A.; Gellerstedt, G. Smooth Model Surfaces from Lignin Derivatives. I. Preparation and Characterization. *Langmuir* **2006**, *22* (3), 1209-1214. DOI: 10.1021/la052284c.

(23) Meyerhofer, D. Characteristics of resist films produced by spinning. *Journal of Applied Physics* **1978**, *49* (7), 3993-3997.

(24) Lawrence, C. The mechanics of spin coating of polymer films. *The Physics of fluids* **1988**, *31* (10), 2786-2795.

(25) Jiang, Y.; Minett, M.; Hazen, E.; Wang, W.; Alvarez, C.; Griffin, J.; Jiang, N.; Chen, W. New insights into spin coating of polymer thin films in both wetting and nonwetting regimes. *Langmuir* **2022**, *38* (41), 12702-12710.

(26) Cousins, W. J. Elastic modulus of lignin as related to moisture content. *Wood Sci Technol* **1976**, *10* (1), 9-17. DOI: 10.1007/BF00376380.

(27) Schwaighofer, M.; Zelaya-Lainez, L.; Königsberger, M.; Lukacevic, M.; Serna-Loaiza, S.; Harasek, M.; Lahayne, O.; Senk, V.; Füssl, J. Characterization of mechanical properties of five hot-pressed lignins extracted from different feedstocks by microscopy-aided nanoindentation. *Mater Design* **2023**, *227*, 111765. DOI: https://doi.org/10.1016/j.matdes.2023.111765.

(28) Sammons, R.; Harper, D.; Labbé, N.; Bozell, J.; Elder, T.; Rials, T. Characterization of Organosolv Lignins using Thermal and FT-IR Spectroscopic Analysis. *Bioresources* **2013**, *8*. DOI: 10.15376/biores.8.2.2752-2767.

(29) Li, W.; Patil, A.; Zhou, X.; Wang, Z.; Xiao, M.; Shawkey, M. D.; Gianneschi, N. C.; Dhinojwala, A. Characterization of broadband complex refractive index of synthetic melanin coatings and their changes after ultraviolet irradiation. *Applied Physics Letters* **2020**, *117* (20).
(30) Hilfiker, J. N.; Stadermann, M.; Sun, J.; Tiwald, T.; Hale, J. S.; Miller, P. E.; Aracne-Ruddle, C. Determining thickness and refractive index from free-standing ultra-thin polymer films with spectroscopic ellipsometry. *Applied Surface Science* **2017**, *421*, 508-512.
(31) Hollertz, R.; Arwin, H.; Faure, B.; Zhang, Y.; Bergström, L.; Wågberg, L. Dielectric properties of lignin and glucomannan as determined by spectroscopic ellipsometry and Lifshitz estimates of non-retarded Hamaker constants. *Cellulose* **2013**, *20* (4), 1639-1648.
(32) Gao, Y.; Ge, K.; Zhang, Z.; Li, Z.; Hu, S.; Ji, H.; Li, M.; Feng, H. Fine optimization of colloidal photonic crystal structural color for physically unclonable multiplex encryption and anti-counterfeiting. *Advanced Science* **2024**, *11* (20), 2305876.
(33) Duan, Z.; Chen, W.; He, Z.; Yang, J.; Cai, G.; Wang, H.; Wang, N.; Hu, Z.; Wang, D.; Chen, N. Reflective angle-insensitive structural colors with a wide color gamut employing a Ni/Ge2Sb2Te5 bilayer broadband absorber. *Optics Express* **2024**, *32* (18), 31866-31879.
(34) Kats, M. A.; Capasso, F. Optical absorbers based on strong interference in ultra-thin films. *Laser & Photonics Reviews* **2016**, *10* (5), 735-749.
(35) Byrnes, S. J. Multilayer optical calculations. *arXiv preprint arXiv:1603.02720* **2016**.
(36) Hecht, E. *Optics*; Pearson Education India, 2012.
(37) Chen, S.; Rossi, S.; Shanker, R.; Cincotti, G.; Gamage, S.; Kühne, P.; Stanishev, V.; Engquist, I.; Berggren, M.; Edberg, J. Tunable structural color images by UV-patterned conducting polymer nanofilms on metal surfaces. *Advanced Materials* **2021**, *33* (33), 2102451.
(38) Rossi, S.; Olsson, O.; Chen, S.; Shanker, R.; Banerjee, D.; Dahlin, A.; Jonsson, M. P. Dynamically tuneable reflective structural coloration with electroactive conducting polymer nanocavities. *Advanced Materials* **2021**, *33* (49), 2105004.
(39) Sardar, S.; Wojcik, P.; Kang, E. S.; Shanker, R.; Jonsson, M. P. Structural coloration by inkjet-printing of optical microcavities and metasurfaces. *Journal of Materials Chemistry C* **2019**, *7* (28), 8698-8704.
(40) Paik, J.; Feng, W.-J.; Clark, S. W.; Kim, H.; Guo, L. J. Iridescent structural color by using ultra-low refractive index aerogel as optical cavity dielectric. *Micro & Nano Manufacturing* **2024**, *1* (1), 1.
(41) Kats, M. A.; Blanchard, R.; Genevet, P.; Capasso, F. Nanometre optical coatings based on strong interference effects in highly absorbing media. *Nature materials* **2013**, *12* (1), 20-24.
(42) Shanker, R.; Mao, A.; Liu, L.; Salhotra, A.; Cui, Y.; An, B.; Jonsson, M. P.; Svagan, A. J. Structurally Colored Thin Films Based on Acetylated Lignin Nanoparticles. *ACS nano* **2025**.
(43) Yang, T.; Mei, C.; Ma, E.; Cao, J. Effects of acetylation on moisture sorption of wood under cyclically changing conditions of relative humidity. *European Journal of Wood and Wood Products* **2023**, *81* (3), 723-731.
(44) Linderoth, O.; Johansson, P.; Wadsö, L. Development of pore structure, moisture sorption and transport properties in fly ash blended cement-based materials. *Construction and Building Materials* **2020**, *261*, 120007.
(45) Voigt, E. M.; Tomlinson, R. The determination of pore size distribution and surface area from adsorption isotherms. *Canadian Journal of Chemistry* **1955**, *33* (2), 215-231.
(46) Wang, J.; Lai, J.; Zhao, W.; Zhou, C.; Hu, C. Preparation of structural colors from lignin: Improving the homogeneity between different raw materials by solvent precipitation fractionation. *Chemical Engineering Journal* **2024**, *499*, 156362.
(47) Chen, J.; Tian, J.; Feng, N.; Ning, L.; Wang, D.; Zhao, B.; Guo, T.; Song, J.; Rojas, O. J. Monodispersed renewable particles by cascade and density gradient size fractionation to advance lignin nanotechnologies. *Small* **2024**, *20* (34), 2309756.
(48) Duval, A.; Vilaplana, F.; Crestini, C.; Lawoko, M. Solvent screening for the fractionation of industrial kraft lignin. In *Holzforschung*, 2016; Vol. 70, p 11.

(49) Thormann, E.; Pettersson, T.; Claesson, P. M. How to measure forces with atomic force microscopy without significant influence from nonlinear optical lever sensitivity. *Review of Scientific Instruments* **2009**, *80* (9).

(50) Hutter, J. L.; Bechhoefer, J. Calibration of atomic-force microscope tips. *Review of scientific instruments* **1993**, *64* (7), 1868-1873.

(51) Hellwig, J.; Durán, V. L.; Pettersson, T. Measuring elasticity of wet cellulose fibres with AFM using indentation and a linearized Hertz model. *Analytical Methods* **2018**, *10* (31), 3820-3823.

(52) Carl, P.; Schillers, H. Elasticity measurement of living cells with an atomic force microscope: data acquisition and processing. *Pflügers Archiv-European Journal of Physiology* **2008**, *457* (2), 551-559.

(53) Senden, T. J. Force microscopy and surface interactions. *Current opinion in colloid & interface science* **2001**, *6* (2), 95-101.

(54) Halbach, K. Matrix representation of Gaussian optics. **1963**.

(55) ElKabbash, M.; Hoffman, N.; Lininger, A. R.; Jalil, S. A.; Letsou, T.; Hinczewski, M.; Strangi, G.; Guo, C. Fano resonant optical coatings platform for full gamut and high purity structural colors. *Nature Communications* **2023**, *14* (1), 3960.